\RequirePackage{fix-cm}
\documentclass[smallextended]{svjour3}       
\smartqed  
\usepackage{graphicx}
\usepackage{amsmath}
\begin{document}

\title{Hadron Spectroscopy: Light, Strange Baryons
}


\author{Chandni Menapara         \and
        Ajay Kumar Rai 
}


\institute{Chandni Menapara         \and
	Ajay Kumar Rai \at
              Department of Physics, Sardar Vallabhbhai National Institute of Technology, Surat-395007, Gujarat, India
              \email{chandni.menapara@gmail.com}           }

\date{Received: date / Accepted: date}

\maketitle

\begin{abstract}
The resonance mass spectra have been studied through a non-relativistic hypercentral Constituent Quark Model (hCQM) using a linear potential. Also, the effects of higher order correction terms (${\cal{O}}(\frac{1}{m})$, ${\cal{O}}(\frac{1}{m^{2}})$) have been studied for improvisation of the results. Other baryonic properties such as Regge trajectories, magnetic moment and decay widths have been considered. A detailed comparison with other approaches are discussed in the present review. 
\end{abstract}

\section{Introduction}\label{sec1}

The current theory of the strong interactions is quantum chromodynamics (QCD), a quantum field theory of quarks and gluons built on the non-abelian gauge group SU \cite{Gross:2022hyw}. Along with the SU(2) $\times$ U(1) electroweak theory, QCD is a part of the Standard Model of particle physics \cite{Leutwyler:2012ax}. QCD is well tested when the strong coupling constant is minimal and perturbation theory is usable at high energies. The strongly coupled QCD theory, of which many aspects are still not fully understood, emerges in the low-energy region. Existing effective degrees of freedom allow us to understand the resonances and bound states of QCD effectively and methodically \cite{Workman:2022ynf}.\\

Confinement serves as QCD's defining characteristic. It appears to forbid isolated fundamental quarks and gluons from existing freely in nature. The gluons' self-interactions, which serve as the strong force's intermediaries for coloured quarks and other gluons, result in confinement. The majority of the mass that is seen in the cosmos is actually created by relativistic interactions between gluons since the light, u, and d quarks that make up the nuclei are a great deal lighter than the proton. The same gluons that hold quarks together and give nucleons their mass can have an impact on the hadron spectrum. Due to the fact that gluons have no charge other than colour, external probes cannot observe them. Fortunately, their contribution to the hadron spectrum is anticipated to result in novel states of matter, and new experimental programs have just started looking for this type of matter, known as glueballs and hybrids, which is dominated by radiation. \\

Over the years, baryons have been explored through various theoretical, phenomenological and experimental aspects \cite{Klempt:2009pi,Rosner_2007,Thiel:2022xtb,Eichmann:2022zxn,Crede:2013kia}. The light, strange baryons from octet and decuplet family comprise of N to $\Omega$ baryon with strangeness S=0 to -3. As it is very well seen from Particle Data Group (PDG)\cite{Workman:2022ynf}, that N, $\Delta$, $\Lambda$, $\Sigma$ baryons have a good number of known states but not all their properties are completely understood, whereas for $\Xi$ and $\Omega$, the states are scarce \cite{Hyodo:2020czb}. The core idea behind this short review is to highlight our predicted resonances alongside the results from other models to provide a concise overview.  The later part also describes some experimental facilities where light sector is being targeted. The third section mentions our methodology and its modifications implemented throughout which is followed by the results and comparison of various models. \\

\subsection{Theoretical Approaches in Light Baryons}
\begin{itemize}

	\item{ \bf Chiral Quark Model}\\
	The success of QCD-inspired models supports the hypothesis being developed by more fundamental studies, according to which QCD is a weakly coupled theory with asymptotically free quark and gluon degrees of freedom below a particular scale, but above this scale, a strong coupling regime emerges in which colour is constrained and chiral symmetry is broken. Nowadays, it is acknowledged that confinement and chiral symmetry breaking are two crucial elements of any QCD-inspired model for the low-energy field. At energies below the scale of spontaneous chiral symmetry breaking (pseudoscalar mesons), it has been suggested that nonstrange and strange baryons can be seen as systems of three quarks interacting via Goldstone boson exchange.  \cite{PhysRevD.81.054002}. Using Chiral Perturbation Theory (ChPT) is a well-known method for analysing hadron structure at low energies.  \\
	
	\item{\bf Relativistic Quark Model}\\
	These models are predicated on quark-diquark hypotheses. To represent baryons as bound states of a constituent quark and diquark, the effective degree of freedom of the diquark is added in quark-diquark models. The structure of baryons is explained using these models. The state space will be drastically reduced since the degrees of freedom of two quarks are locked in place within the diquark. \cite{Ebert:2011kk}. Diquarks in strange baryons are formed by the constituent quarks of the same mass as ud, uu, dd and ss. Also the ground state ud diquark can be both in scalar and axial vector state, while the ground state diquarks composed from quarks of the same flavour uu, dd and ss can be only in the axial vector state due to the Pauli principle \cite{Faustov:2015eba}. One such approach includes the the substitution of the spin and isospin dependent terms by Gursey and Radicati-inspired exchange interaction. \\
	
	\item{ \bf Lattice QCD}\\
	Lattice QCD calculations of hadron spectroscopy have made impressive strides towards large-volume simulations with physical quark masses, impressive statistical precision, and good control of different systematic uncertainties \cite{Gupta:1997nd}.
	Multiple lattice groups have been conducting in-depth, systematic studies of stable hadrons that are far below the minimum allowed strong decay threshold.
	Lattice QCD is quantized in accordance with Feynman's path integral formalism, and it is a gauge field theory that operates on a 4D Euclidean lattice space-time.
	Reliable computations of the simplest hadronic matrix elements required to extract CKM quark-mixing matrix elements and for many Standard-Model tests are already being provided by lattice QCD. The parameters of QCD can then be calculated with great ease and precision using lattice QCD \cite{UKQCD:1996ssj}. Lattice computations have been essential in the measurement of the properties of quarks, and they now provide the quantitative predictions of hadronic matrix elements needed for flavour physics, allowing for the first time ever the computation of the spectrum of hadrons with high accuracy \cite{Edwards:2012fx,Mathur:2018epb}.  Recently some of the states  of $\Lambda$ baryon have been studied away from SU(3) limit \cite{Guo:2023wes}\\
	
	\item{ \bf QCD Sum Rules}\\
	Instead of a model-dependent approach in terms of constituent quarks, hadrons are represented by their interpolating quark currents taken at large virtualities.
	The operator product expansion (OPE), which distinguishes between short- and long-range quark-gluon interactions, is the framework in which the correlation function of these currents is presented and discussed. While the latter are specified by universal hoover condensates or light-cone distribution amplitudes, the former are computed using QCD perturbation theory.
	A dispersion relation is then used to match the QCD calculation's output to a sum over hadronic states. The calculation of observable hadronic ground state attributes is made possible by the sum rule established in this way, as noted in \cite{Dominguez:2013ata}. So, Shifman, Vainshtein, and Zakharov's method of QCD sum rules, developed over twenty years ago, has become a widely used working tool in hadron phenomenology \cite{Colangelo:2000dp}.\\
	
	\item{ \bf Regge Phenomenology}\\
	With the application of Regge's theory to the study of strong interaction, Chew and Frautschi discovered that hadrons can be located on the linear trajectories of the $(J,M^{2})$ plane. The path is the same for hadrons with the same internal quantum numbers. The linearity of the Regge trajectory may be explained by the idea that quarks and antiquarks were connected to one another by a gluon flux tube. Along the radius, the light quarks at the rotating mass's ends move at the speed of light. Regge phenomenology has been intensively explored to obtain the resonance mass through using the slope and intercept of the linear nature of J and n- plane trajectories \cite{Guo:2008he}. Recently, our group has also explored a sector of heavy, light systems under the light of Regge phenomenology through slope and intercept equations \cite{Oudichhya:2021yln,Oudichhya:2021kop}.\\
	
	\item{ \bf Relativistic Flux-Tube Model}\\
	Nambu's relativistic flux tube model (RFT) is one of the simplest explanations for the linear Regge trajectory \cite{nambu1974strings}. The RFT model assumes, at its most fundamental level, that the strong interaction is mediated by a rigid and straight tube-like structure made up of gluonic pitch only, connecting the quarks within hadrons \cite{Jakhad:2023mni}. It is assumed that the whole system is spinning around its own mass centre.\\
	
	\item{ \bf Other theoretical models}\\
	In the late 1960s and early 1970s, the Bogoliubov model served as the foundation for what became known as the Bag model. The hadronic states can maintain reasonable masses by considering quarks to be massive enough and bound in a deep potential. For his proposal, it is assumed that the quark masses are m $\rightarrow$ $\infty$ and confined them to a sphere of radius R  where they experienced an attractive scalar field.\\
	The Skyrme model, which deals with topological solitons, is a popular variant of the soliton models. The non-linear sigma model's effective Lagrangian forms the basis of this model \cite{chen2020bottomonium,LaCourse:1988cu}. Recent times, light-front holography approach has also been of keen interest \cite{Brodsky:2021dze}. \\
	There are varied models in addition to the aforementioned methods. Algebraic model by Bijker et al. \cite{Bijker:1998zb} studies the observable quantities of hadronic systems by collective string-like model. An extension to this is obtaining the algebraic solution by Bethe Ansatz within an infinite-dimensional Lie algebra U(7) \cite{amiri2021transitional}. The 1/Nc expansion approach for excited baryons has been implemented under the assumption that there is an approximate spin-flavour symmetry in the large Nc limit \cite{PhysRevD.66.114014}. Chen and group has shown a simple classification of baryons based on mass range in n and $J^{P}$ values. E. Klempt has given an important insight into baryon spectra studies through mass formula \cite{KLEMPT20071}. A study using Poincare covariant quark-diquark Faddeev approach has been applied towards exploring the structure of light baryon multiplets \cite{PhysRevD.105.114047}. \\
	Phenomenological models give us a window into the structure of hadrons because they retain and make use of key features of QCD.
\end{itemize}

\subsection{Experiments in Light Baryon Spectroscopy}

Here, we focus the ongoing and upcoming experimental facilities particularly targeted for or around light, strange sector.
In Europe, there is an international accelerator facility called the Facility for Antiproton and Ion Research (FAIR) where scientists can study the properties of antiprotons and ions. 

The Hades spectrometer is a versatile detector device operating at the SIS18 synchrotron at GSI Darmstadt with a vital list of results on strangeness production including $\Lambda$(1405) and $\Sigma$(1385) exclusive production cross-sections \cite{HADES:2020pcx}.
\begin{itemize}

	\item {\bf $\bar{P}$ANDA}
	
	The $\bar{P}$ANDA experiment is one of the major projects of FAIR. Since the PANDA detector is located at the HESR (High Energy Storage Rings), studies of interactions between the antiproton beam and fixed target protons and nuclei are
	performed \cite{Frankfurt:2018msx}. The antiprotons will have momenta in the range of 1.5 GeV/c and 15 GeV/c. Mostly in the non-perturbative regime, the scientific program seeks to provide answers to the fundamental questions of QCD. $\bar{p}p$ collisions enable all non-exotic quantum number combinations for directly formed states, which is complementary to $e^{+}e^{-}$ induced reactions. \\
	
	PANDAs environment to produce abundant pairs of hyperons and antihyperon is the ideal setting to carry out detailed spectroscopy studies of these baryons \cite{PANDA:2021ozp}. In the case of PANDA, the conceptual idea is to replace light valence quarks of the (anti)proton
	with heavier strange and charm ones, measure the excitation spectrum of excited
	hyperon states, determine their properties such as mass, width, spin, parity, and decay modes, and compare such observations between the various baryonic
	systems including those of the light-quark sector, i.e. N and $\Delta$ resonance levels. The PANDA detector will give the possibility to determine the electromagnetic form factors of the Proton in the time-like region with high
	accuracy \cite{PANDA:2020jkm,PANDA:2017sxm}. Due to the possibility of a precise multi-strange hyper-nucleus spectroscopy at PANDA, it will be possible to explore the hyperon-hyperon interaction \cite{PANDA:2021ozp,Dbeyssi:2022zwz,PANDA:2023ljx,2016323}.
	
	\item {\bf BGO-OD at ELSA}
	The BGO-OD experiment at the ELSA accelerator facility uses an energy-tagged bremsstrahlung photon beam to investigate the excitation structure of the nucleon. A highly segmented BGO calorimeter is around the target, and a particle tracking magnetic spectrometer is positioned at forward angles. \cite{BGO-OD:2019utx,Scheluchin:2020jhr}. The extensive strangeness photo production programme also uses hydrogen and deuterium targets to produce neutral and charged kaons.
	
	\item {\bf J-PARC}
	The Japan Proton Accelerator Research Complex (J-PARC) is a multi-purpose accelerator facility located in Tokai, Japan \cite{Aoki:2021cqa}. The mission of J-PARC is to advance a wide range of scientific research initiatives, from the fundamentals of particle, nuclear, atomic, and condensed matter physics to the application of powerful particle beams in industry and nuclear transmutation in the future.
	At the energy of J-PARC, excited baryons with a charm quark or multi strange
	quarks are appropriate. The J-PARC accelerator consists of a 400-MeV linac
	as an injector, a 3-GeV rapid-cycling synchrotron (RCS), and the 30 GeV main-ring synchrotron. Spectroscopy of $\Xi$ baryons is planned at the new K10 with intense, separated kaons to investigate the diquark correlation in the strangeness sector \cite{Ohnishi:2019cif}.
	
	\item {\bf J-Lab}
	The JLab 12 GeV energy upgrade, with the new Hall D, provides an ideal tool for extensive studies of both non-strange and, specifically, strange baryon resonances \cite{Dudek:2012vr}.The principal goal is to provide large
	acceptance at high luminosity so that small cross sections can be measured with high precision. Many hyperon spectroscopy measurements are expected from
	the GlueX and CLAS12 measurements, including the $\Xi$ and $\Omega$ \cite{Arrington:2021alx}. This program will be expanded to perform hyperon spectroscopy with the KL neutral kaon beam in Hall D \cite{KLF:2020gai}.
	
	\item {\bf BESIII}
	At the Beijing Electron Positron Collider (BEPCII), the BESIII collaboration uses centre-of-mass $e^+e^-$ collisions to measure the properties of heavy ions \cite{BESIII:2020nme,BESIII:2022mxl}.
	The centre-of-mass energies between 2.0 and 5.0 GeV investigates the vast physics landscape available at these energies. BESIII has so far collected 40 fb- of data since 2009. All SU(3) octet hyperons and several charmed baryons have their production thresholds within BESIII's energy region.
	Born cross sections for electron-positron annihilation to different baryon pairs, such as $\Lambda\bar{\Lambda}$, $\Sigma\bar{\Sigma}$, $\Xi\bar{\Xi}$, are measured from threshold at BESIII \cite{BESIII:2019dve}.\\

	\item {\bf NICA Program}
	A project at the Joint Institute for Nuclear
	Research; The Nuclotron-based Ion Collider fAcility (NICA) is
	a new research complex for studying the fundamental
	properties of the strong interaction being developed \cite{Kekelidze:2016hhw}. Collisions
	of high-intensity proton beams with a high degree
	of longitudinal or transverse polarization and with total
	energy up to 13.5 GeV will be there. Using ion beams scavenged from the modernized Nuclotron, BM@N is a fixed-target experimental setup.
	Through the registration of strange and multi-strange particle (kaons, and hyperons, double hypernuclei, etc.) production, the BM@N experiment seeks to comprehensively study the early phase of nuclear interaction at high densities of nuclear matter \cite{Steinheimer:2009zza}. Strangeness abounds in heavy ion collisions, and capturing $\Lambda$-hyperons by nucleons can yield several different types of light hyper-nuclei.
	
	\item {\bf SLAC-BABAR}
	To better understand CP-Violation in the decay of B mesons, the B-factory Experiment at SLAC was designed. Large numbers of charm particles are created as a by-product, and charm baryons in particular that decay to final states containing hyperons and hyperon resonances are reconstructed at a reasonable statistical level \cite{BaBar:2005uwk}. The accelerator has asymmetric beam energies, with a 9 GeV electron beam colliding head on with a 3.1 GeV positron beam.
	
	\item {\bf LHC}
	The ALICE Collaboration at Large Hadron Collider (LHC) observed a proton and $\Xi$ interaction. In proton-proton (pp) and proton-lead (p-Pb) collisions at the LHC, hyperon-nucleon, hyperon-hyperon  and kaon-nucleon interactions can be measured with greater precision \cite{ALICE:2019hdt}. In fact, at LHC energies, small colliding systems produce particle-emitting sources that are about 1 fm in size, enabling a precise test of the short-range strong interaction. These studies can be expanded to include the p-$\Xi$ correlation thanks to the larger number of pairs present in the data set obtained from p-Pb collisions at  $\sqrt{sNN}$ = 5.02TeV by ALICE, which have an emitting source size similar to that of pp collisions \cite{PhysRevLett.118.052002}.\\
	These studies are also important for neutron star modelling because neutrons may become hyperons to reduce system energy because of the high densities attained in the centres of these objects.
	
	\item {\bf CB-ELSA/TAPES}
	The $N\pi\pi$ decays of positive-parity N and $\Delta$ resonances at about 2 GeV are studied at ELectron Stretcher Accelerator (ELSA) at the University of Bonn, and the Crystal Barrel detector by photo-production of two neutral pions off protons i.e. $\gamma p \rightarrow p \pi^0 \pi^0$ \cite{Thoma:2007bm,Hillert}. \\
\end{itemize}
\begin{table*}
	
	\caption{\label{tab:nucleon} Nucleon mass spectra (in MeV)}
	\begin{tabular}{@{\extracolsep{\fill}}cccccccccccccccc}
		
		\hline
		State &  $J^{P}$ &  $M_{cal}$  & PDG\cite{Workman:2022ynf} & Status & \cite{Santopinto:2014opa} & \cite{amiri2021transitional} & \cite{Anisovich:2011fc} & \cite{Ghalenovi:2017xxv} & \cite{Chen:2009de} & \cite{Aslanzadeh:2017hhz} & \cite{Klempt:2002vp} \\
		\hline
		1S & $\frac{1}{2}^{+}$ &  939 & 938 & **** & 939 & 939 & 960 & 938 & 939 & 938 & 938 \\ 
		2S & $\frac{1}{2}^{+}$ &  1440 & 1410-1470 & **** & 1511 & 1450.8 & 1430 & 1444 & 1462 & 1492 & 1440  \\
		3S & $\frac{1}{2}^{+}$ &  1737 & 1680-1740 & *** & 1776 & 1699 & 1710 & 1832 & 1748 & 1763 & 1710  \\
		4S & $\frac{1}{2}^{+}$ & 2034 & 2050-2150 & *** & & & & & & &  \\
		\hline
		1P & $\frac{1}{2}^{-}$ &  1519 &  1515-1545 & **** & 1537 & 1536.1 & 1501 & 1567 & 1497 & 1511 & 1538 \\
		1P & $\frac{3}{2}^{-}$ & 1537 &  1510-1520 & **** & 1537 & 1550.3 & 1517 & 1567 & 1548 & 1511 & 1523  \\
		1P & $\frac{5}{2}^{-}$ & 1542 & 1665-1680 & **** & & & & & 1655 & & 1678 \\
		\hline
		1D & $\frac{1}{2}^{+}$ &  1750 & 1830-1930 & *** & 1890 & & 1870 & 1887 & & & \\
		1D & $\frac{3}{2}^{+}$ & 1761 & 1660-1750 & **** & 1648 & 1682.7 & 1690 & & 1734 & 1725 & 1700  \\
		1D & $\frac{5}{2}^{+}$ &  1773 & 1680-1690 & **** & 1799 & 1704.2 & 1689 & 1689.8 & 1738 & 1735 & 1683  \\
		1D & $\frac{7}{2}^{+}$ & 1781 &  & & & & & & 1943 & & 1990 \\
		\hline
		1F & $\frac{3}{2}^{-}$ & 2033 & 2060-2160 & *** & & & & & & & 2080  \\
		1F & $\frac{5}{2}^{-}$ &   2045 & 2030-2200 & *** & & & & & & & 2220  \\
		1F & $\frac{7}{2}^{-}$ &  2058 &  2140-2220 & **** & & 2135 & 2180 & & &  & 2150 & \\
		1F & $\frac{9}{2}^{-}$ &  2068 & 2250-2320 & **** & & 2270 & 2280 & 2232.4 &  & 2240 \\
		\hline
		1G & $\frac{9}{2}^{+}$ & 2374 & 2200-2300 & **** & & 2273 & 2200 & 2174.3 & & & 2245 \\
		\hline
		1H & $\frac{11}{2}^{-}$ & 2713 & 2550-2750 & *** & & 2620 & & 2534.5 & & & 2650  \\
		\hline
	\end{tabular}
\end{table*}
\begin{table}
	\caption{\label{tab:lambda} $\Lambda$ mass spectra compared to results from Regge phenomenology and relativistic model (in MeV)} 
	\begin{tabular}{ccccccc}
		\hline
		State & $J^{P}$ & $M_{cal}$\cite{Menapara:2023rur} & $M_{exp}$\cite{Workman:2022ynf} & \cite{OUDICHHYA2023122658} & \cite{Faustov:2015eba} & \\
		\hline
		1S & $\frac{1}{2}^{+}$ &  1115 &  1115 & 1115 & 1115\\
		2S & $\frac{1}{2}^{+}$ &  1589 &  1600 & 1600 & 1615 \\
		3S & $\frac{1}{2}^{+}$ & 1892 &  1810 & 1969 & 2099 \\
		4S & $\frac{1}{2}^{+}$ &  2220 &  & 2278 & 2546\\
		5S & $\frac{1}{2}^{+}$ & 2571 & & 2551 & &\\
		\hline
		$1^{2}P_{1/2}$ & $\frac{1}{2}^{-}$ & 1535 & & & \\
		$1^{2}P_{3/2}$ & $\frac{3}{2}^{-}$ & 1540 &  1520 & 1551 & 1549\\
		$1^{4}P_{1/2}$ & $\frac{1}{2}^{-}$ &  1533 & 1670 & & 1667\\
		$1^{4}P_{3/2}$ & $\frac{3}{2}^{-}$ &  1542 & 1690 & & 1693 \\
		$1^{4}P_{5/2}$ & $\frac{5}{2}^{-}$ & 1543 & & & &\\
		\hline
		$2^{2}P_{1/2}$ & $\frac{1}{2}^{-}$ & 1831 & 1800 & & \\
		$2^{2}P_{3/2}$ & $\frac{3}{2}^{-}$ & 1837 &  & 1690 & 1812 \\
		$2^{4}P_{1/2}$ & $\frac{1}{2}^{-}$ & 1829 & \\
		$2^{4}P_{3/2}$ & $\frac{3}{2}^{-}$ & 1837 & \\
		$2^{4}P_{5/2}$ & $\frac{5}{2}^{-}$ & 1840 & 1830 \\
		\hline
		$1^{2}D_{3/2}$ & $\frac{3}{2}^{+}$ & 1762 & \\
		$1^{2}D_{5/2}$ & $\frac{5}{2}^{+}$ &  1766 & 1820 & 1889 & 1825\\
		$1^{4}D_{1/2}$ & $\frac{1}{2}^{+}$ & 1756 & 1710 \\ 
		$1^{4}D_{3/2}$ & $\frac{3}{2}^{+}$ & 1761 & 1890\\
		$1^{4}D_{5/2}$ & $\frac{5}{2}^{+}$ & 1766 & \\
		$1^{4}D_{7/2}$ & $\frac{7}{2}^{+}$ & 1768 & \\
		\hline
		$1^{2}F_{5/2}$ & $\frac{5}{2}^{-}$ &  1999 & 2080 & & 2136 \\
		$1^{2}F_{7/2}$ & $\frac{7}{2}^{-}$ &  2004 & 2100 & 2175 & 2097\\
		$1^{4}F_{3/2}$ & $\frac{3}{2}^{-}$ &  1993 & \\
		$1^{4}F_{5/2}$ & $\frac{5}{2}^{-}$ &  1999 & \\
		$1^{4}F_{7/2}$ & $\frac{7}{2}^{-}$ &  2004 & \\
		$1^{4}F_{9/2}$ & $\frac{9}{2}^{-}$ &  2008 & \\
		\hline
		$1^{2}G_{7/2}$ & $\frac{7}{2}^{+}$ & 2248 & & & 2251\\
		$1^{2}G_{9/2}$ & $\frac{9}{2}^{+}$ & 2255 & 2350 & 2427 & 2360 \\
		$1^{4}G_{5/2}$ & $\frac{5}{2}^{+}$ & 2239 & \\
		$1^{4}G_{7/2}$ & $\frac{7}{2}^{+}$ & 2247 & \\
		$1^{4}G_{9/2}$ & $\frac{9}{2}^{+}$ & 2254 & \\
		$1^{4}G_{11/2}$ & $\frac{11}{2}^{+}$ & 2260 & \\
		\hline
	\end{tabular}
\end{table}

\begin{table}
	\caption{\label{tab:Sigma} $\Sigma$ mass spectra compared to results from Regge phenomenology and relativistic model (in MeV)} 
	\begin{tabular}{cccccc}
		
		\hline
		State & $J^{P}$ & $M_{cal}$\cite{Menapara:2023rur} & $M_{exp}$\cite{Workman:2022ynf} & \cite{OUDICHHYA2023122658} & \cite{Faustov:2015eba}\\
		\hline
		1S & $\frac{1}{2}^{+}$ & 1193 & 1193 & 1193 & 1187\\
		& $\frac{3}{2}^{+}$ & 1384 &  1385 & 1384 & 1381\\
		2S & $\frac{1}{2}^{+}$ & 1643 & 1660 & 1711 & 1711\\
		& $\frac{3}{2}^{+}$ & 1827 & 1780* & 1862 & 1862\\
		3S & $\frac{1}{2}^{+}$ & 2099 & & 2105 & 2292 \\
		& $\frac{3}{2}^{+}$ & 2236 & 2230*/2250 & 2240 & 2347\\
		4S & $\frac{1}{2}^{+}$ & 2589 & 2620* & 2437 & 2740 \\
		& $\frac{3}{2}^{+}$ & 2693 & & 2564 & \\ 
		5S & $\frac{1}{2}^{+}$ & 3108 & 3000* & &\\
		& $\frac{3}{2}^{+}$ & 3189 & 3170* & &\\
		\hline
		$1^{2}P_{1/2}$ & $\frac{1}{2}^{-}$ & 1687 & 1620* & & 1620 \\
		$1^{2}P_{3/2}$ & $\frac{3}{2}^{-}$ & 1696 & 1670 & 1607 & 1706\\
		$1^{4}P_{1/2}$ & $\frac{1}{2}^{-}$ & 1684 & 1750 & & 1693 \\
		$1^{4}P_{3/2}$ & $\frac{3}{2}^{-}$ & 1698 & & & 1731 \\
		$1^{4}P_{5/2}$ & $\frac{5}{2}^{-}$ & 1700 & 1775 & 1771 & 1757 \\
		\hline
		$2^{2}P_{1/2}$ & $\frac{1}{2}^{-}$ & 2098 & 1900 & & 2115 \\
		$2^{2}P_{3/2}$ & $\frac{3}{2}^{-}$ &  2106 & 1910 & 2175 & 2175\\
		$2^{4}P_{1/2}$ & $\frac{1}{2}^{-}$ & 2095 & 2110* & & 2198\\
		$2^{4}P_{3/2}$ & $\frac{3}{2}^{-}$ & 2106 & 2010* & & 2203 \\
		$2^{4}P_{5/2}$ & $\frac{5}{2}^{-}$ & 2110 & & 2214 & 2214\\
		
		\hline
		$1^{2}D_{3/2}$ & $\frac{3}{2}^{+}$ &  2003 & 1940* & & 2025\\
		$1^{2}D_{5/2}$ & $\frac{5}{2}^{+}$ &  2010 & 1915 & 1935 & 1991 \\
		$1^{4}D_{1/2}$ & $\frac{1}{2}^{+}$ & 1992 & 1880* & & 1922 \\
		$1^{4}D_{3/2}$ & $\frac{3}{2}^{+}$ &  2002 & 2080* & & 2076\\
		$1^{4}D_{5/2}$ & $\frac{5}{2}^{+}$ &  2010 & 2070* & & 2062\\
		$1^{4}D_{7/2}$ & $\frac{7}{2}^{+}$ & 2015 & 2025 & 2087 & 2033 \\
		\hline
		$2^{2}D_{3/2}$ & $\frac{3}{2}^{+}$ & 2448 & 2455* & & 2465\\
		$2^{2}D_{5/2}$ & $\frac{5}{2}^{+}$ & 2456 & & 2459 & 2459 \\
		$2^{4}D_{1/2}$ & $\frac{1}{2}^{+}$ & 2438 & & & 2472\\
		$2^{4}D_{3/2}$ & $\frac{3}{2}^{+}$ &  2446 & & & 2465\\
		$2^{4}D_{5/2}$ & $\frac{5}{2}^{+}$ &  2455 & & & 2459\\
		$2^{4}D_{7/2}$ & $\frac{7}{2}^{+}$ & 2462 & 2470 & & 2470\\
		\hline
		$1^{2}F_{5/2}$ & $\frac{5}{2}^{-}$ & 2339 & & & 2347\\
		$1^{2}F_{7/2}$ & $\frac{7}{2}^{-}$ &  2349 & 2100* & 2215 & 2259 \\
		$1^{4}F_{3/2}$ & $\frac{3}{2}^{-}$ & 2327 &  & 2333 & 2409 \\
		$1^{4}F_{5/2}$ & $\frac{5}{2}^{-}$ &  2337 & & & 2459 \\
		$1^{4}F_{7/2}$ & $\frac{7}{2}^{-}$ & 2348 & & 2249 & 2349 \\
		$1^{4}F_{9/2}$ & $\frac{9}{2}^{-}$ & 2356 & & 2362 & 2289 \\
		\hline
		
	\end{tabular}
\end{table}

\begin{table}
	\caption{\label{tab:XI} $\Xi$ mass spectra compared to results from Regge phenomenology and relativistic model (in MeV)}
	\begin{tabular}{cccccc}
		
		\hline
		State & $J^{P}$ & $M_{cal}$\cite{Menapara:2023rur} & $M_{exp}$ \cite{Workman:2022ynf} & \cite{OUDICHHYA2023122658} & \cite{Faustov:2015eba} \\
		\hline
		1S & $\frac{1}{2}^{+}$ & 1321 & 1321 & 1292 & 1330\\
		& $\frac{3}{2}^{+}$ & 1524 &  1532 & 1534 & 1518\\
		2S & $\frac{1}{2}^{+}$ & 1891 &  & 1886 & 1886\\
		& $\frac{3}{2}^{+}$ & 1964 & 1950 & 1966 & 1966\\
		3S & $\frac{1}{2}^{+}$ & 2372  & 2370 & 2334 & 2367\\
		& $\frac{3}{2}^{+}$ & 2459 & & 2319 & 2421 \\
		4S & $\frac{1}{2}^{+}$ & 2954 & & 2708 &  \\
		& $\frac{3}{2}^{+}$ & 3041 &  & 2624 & \\ 
		5S & $\frac{1}{2}^{+}$ & 3620 & & \\
		& $\frac{3}{2}^{+}$ & 3702 & & \\
		\hline
		$1^{2}P_{1/2}$ & $\frac{1}{2}^{-}$ & 1862 & & 1810 & 1758 \\
		$1^{2}P_{3/2}$ & $\frac{3}{2}^{-}$ & 1869 & 1823 & 1731 & 1764\\
		$1^{4}P_{1/2}$ & $\frac{1}{2}^{-}$ & 1860 & & 1890 & \\
		$1^{4}P_{3/2}$ & $\frac{3}{2}^{-}$ & 1870 & 1823 &  1825 & 1798\\
		$1^{4}P_{5/2}$ & $\frac{5}{2}^{-}$ & 1872 & & 1923 & 1853 \\
		\hline
		$2^{2}P_{1/2}$ & $\frac{1}{2}^{-}$ & 2333 & 2370 & & 2160\\
		$2^{2}P_{3/2}$ & $\frac{3}{2}^{-}$ & 2340 & & 2245 & 2245\\
		$2^{4}P_{1/2}$ & $\frac{1}{2}^{-}$ & 2330 & & 2233\\
		$2^{4}P_{3/2}$ & $\frac{3}{2}^{-}$ & 2341 & & 2252\\
		$2^{4}P_{5/2}$ & $\frac{5}{2}^{-}$ & 2344 & & 2333 & 2333\\
		\hline
		
		$1^{2}D_{3/2}$ & $\frac{3}{2}^{+}$ & 2236 &  & 2155 & \\
		$1^{2}D_{5/2}$ & $\frac{5}{2}^{+}$ & 2241 & 2250 & 2080 & \\
		$1^{4}D_{1/2}$ & $\frac{1}{2}^{+}$ & 2228 & & \\
		$1^{4}D_{3/2}$ & $\frac{3}{2}^{+}$ & 2235 & & 2204 &\\
		$1^{4}D_{5/2}$ & $\frac{5}{2}^{+}$ & 2242 & & 2115 &\\
		$1^{4}D_{7/2}$ & $\frac{7}{2}^{+}$ & 2245 & & 2245 & \\
		\hline
		$2^{2}D_{3/2}$ & $\frac{3}{2}^{+}$ & 2782 & & \\
		$2^{2}D_{5/2}$ & $\frac{5}{2}^{+}$ & 2788 & & 2605 &\\
		$2^{4}D_{1/2}$ & $\frac{1}{2}^{+}$ & 2772 & & \\
		$2^{4}D_{3/2}$ & $\frac{3}{2}^{+}$ & 2780 & & \\
		$2^{4}D_{5/2}$ & $\frac{5}{2}^{+}$ & 2788 & & \\
		$2^{4}D_{7/2}$ & $\frac{7}{2}^{+}$ & 2793 & & 2686 & \\
		\hline
		$1^{2}F_{5/2}$ & $\frac{5}{2}^{-}$ & 2664 &  & 2318 & 2400\\
		$1^{2}F_{7/2}$ & $\frac{7}{2}^{-}$ & 2671 & & & 2460\\
		$1^{4}F_{3/2}$ & $\frac{3}{2}^{-}$ & 2654 & & & \\
		$1^{4}F_{5/2}$ & $\frac{5}{2}^{-}$ & 2662 & & & 2455\\
		$1^{4}F_{7/2}$ & $\frac{7}{2}^{-}$ & 2670 & & 2370 & 2474\\
		$1^{4}F_{9/2}$ & $\frac{9}{2}^{-}$ & 2676 & & & 2502\\
		\hline
		
	\end{tabular}
\end{table}

\begin{table}
	\caption{\label{tab:Omega} $\Omega$ mass spectra compared to results from Regge phenomenology and relativistic model (in MeV) } 
	\begin{tabular}{cccccc} 
		\hline
		State & $J^{P}$ &  $M_{cal}$\cite{Menapara:2023rur} & $M_{exp}$\cite{Workman:2022ynf} & \cite{OUDICHHYA2023122658} & \cite{Faustov:2015eba} \\
		\hline
		1S & $\frac{3}{2}^{+}$ & 1672  &  1672 & 1686 & 1678 \\
		2S & $\frac{3}{2}^{+}$ & 2068  & & 2057 & 2173\\
		3S & $\frac{3}{2}^{+}$ & 2449  & & 2371 & 2671\\
		4S & $\frac{3}{2}^{+}$ & 2885  & & 2647 & \\
		\hline
		$1^{2}P_{1/2}$ & $\frac{1}{2}^{-}$ & 1979 & \\
		$1^{2}P_{3/2}$ & $\frac{3}{2}^{-}$ & 1983  \\
		$1^{4}P_{1/2}$ & $\frac{1}{2}^{-}$ & 1977 & 2012 & 2030 & 1941\\
		$1^{4}P_{3/2}$ & $\frac{3}{2}^{-}$ & 1984 & & 1941 & 2038\\
		$1^{4}P_{5/2}$ & $\frac{5}{2}^{-}$ & 1985 & & 2078 & \\
		\hline
		$2^{2}P_{1/2}$ & $\frac{1}{2}^{-}$ & 2342 & 2380 & \\
		$2^{2}P_{3/2}$ & $\frac{3}{2}^{-}$ & 2346 \\
		$2^{4}P_{1/2}$ & $\frac{1}{2}^{-}$ & 2340 & & & 2463\\
		$2^{4}P_{3/2}$ & $\frac{3}{2}^{-}$ & 2346 & & & 2537\\
		$2^{4}P_{5/2}$ & $\frac{5}{2}^{-}$ & 2348 & & 2321 &\\
		\hline
		
		$1^{2}D_{3/2}$ & $\frac{3}{2}^{+}$ & 2263 & 2250 \\
		$1^{2}D_{5/2}$ & $\frac{5}{2}^{+}$ & 2266 \\
		$1^{4}D_{1/2}$ & $\frac{1}{2}^{+}$ & 2258 \\
		$1^{4}D_{3/2}$ & $\frac{3}{2}^{+}$ & 2263 & & 2338 & 2332 \\
		$1^{4}D_{5/2}$ & $\frac{5}{2}^{+}$ & 2266 & & 2249 & 2401 \\
		$1^{4}D_{7/2}$ & $\frac{7}{2}^{+}$ & 2267 & & 2407 & 2369 \\
		\hline
		$2^{2}D_{3/2}$ & $\frac{3}{2}^{+}$ & 2671 \\
		$2^{2}D_{5/2}$ & $\frac{5}{2}^{+}$ & 2674 \\
		$2^{4}D_{1/2}$ & $\frac{1}{2}^{+}$ & 2666 \\
		$2^{4}D_{3/2}$ & $\frac{3}{2}^{+}$ & 2670 \\
		$2^{4}D_{5/2}$ & $\frac{5}{2}^{+}$ & 2674 \\
		$2^{4}D_{7/2}$ & $\frac{7}{2}^{+}$ & 2676  & & 2623\\
		\hline
		$1^{2}F_{5/2}$ & $\frac{5}{2}^{-}$ & 2578 \\
		$1^{2}F_{7/2}$ & $\frac{7}{2}^{-}$ & 2581 \\
		$1^{4}F_{3/2}$ & $\frac{3}{2}^{-}$ & 2573 \\
		$1^{4}F_{5/2}$ & $\frac{5}{2}^{-}$ & 2577 & & 2610 & 2653\\
		$1^{4}F_{7/2}$ & $\frac{7}{2}^{-}$ & 2582 & & 2519 & 2599\\
		$1^{4}F_{9/2}$ & $\frac{9}{2}^{-}$ & 2584 & & & 2649 \\
		\hline
		
	\end{tabular}
\end{table}
\section{Methodology: hCQM}
Our current understanding of hadrons is based on the success of QCD and the spectroscopic description of hadrons in terms of more massive constituent quarks. For instance, the three valence quarks of the nucleon are dressed with gluons and quark antiquark pairs to become noticeably heavier. \\

It is possible to assess the baryon properties starting from a quark model if any interaction between the constituent quarks is added. One must rely on the fundamental prerequisites because there hasn't been any hint of the shape of this interaction potential up to this time. The quark-quark interaction is known to be necessary.

A baryon or meson is modelled as a system of three quarks, anti-quarks, or quark-antiquark pairs bound by some sort of confining interaction in a constituent quark model \cite{Ferraris:1995ui}. With this straightforward presumption, a model that can provide a quantitative description of a number of hadron properties can be constructed. With the addition of a large constituent quark mass, all effects in a CQM that go beyond the description of a baryon as a system of three confined particles such as the effects of the gluons and sea quarks are effectively parametrized.
In a theory known as the constituent quark model, baryons are thought of as a bound, colourless system made up of three particles known as the constituent quarks \cite{Giannini:2001kb}. The effective degrees of freedom of the three constituent quarks, which are fermionic particles with flavour (SU(3)flavour) and colour (SU(3)colour) degrees of freedom, serve as the foundation for all of the CQMs. 
The three quark wave-function can be written as:
\begin{equation}
	\psi = \phi_{flavour} \chi_{spin} \xi_{colour} \eta_{space} 
\end{equation}
It is generally accepted that the Hamiltonian operator for three quarks takes the form H = K + V, where K refers to the non-relativistic kinetic energy operator and V refers to the three-quark interaction. Because of the non-abelian nature of QCD, which results in gluon-gluon interactions, which, in turn, can produce three body forces, these terms have the potential to play an important part in the description of hadrons.

The hypercentral Constituent Quark Model, abbreviated as hCQM, is predicated on the selection of a hypercentral SU(6) invariant potential, also known as a potential whose value is determined solely by the hyperradius x \cite{Shah:2023zph,Kakadiya:2021jtv}. In addition to making the process of finding a solution for the Schr\"{o}edinger equation more straightforward, the selection of a hypercentral potential also has some intriguing ramifications in terms of the physical world \cite{Patel:2007gx,Shah:2016mig}. {\bf One such study has been carried out using artificial neural network by H. Mutuk \cite{Mutuk:2020rzm,Mutuk:2021epz}. }
\begin{equation}
	{\bf \rho} = \frac{1}{\sqrt{2}}({\bf r_{1}} -{\bf r_{2}}); \; \; {\bf \lambda} = \frac{1}{\sqrt{6}} ( {\bf r_{1}} + {\bf r_{2}} - 2{\bf r_{3}})
\end{equation} 
\begin{figure}
	\centering
	
	\caption{\label{fig:radial}Tentative depiction of quark during radial excitation in symmetry and mixed-symmetry condition \cite{Capstick:2000qp}}
	\includegraphics[scale=0.5]{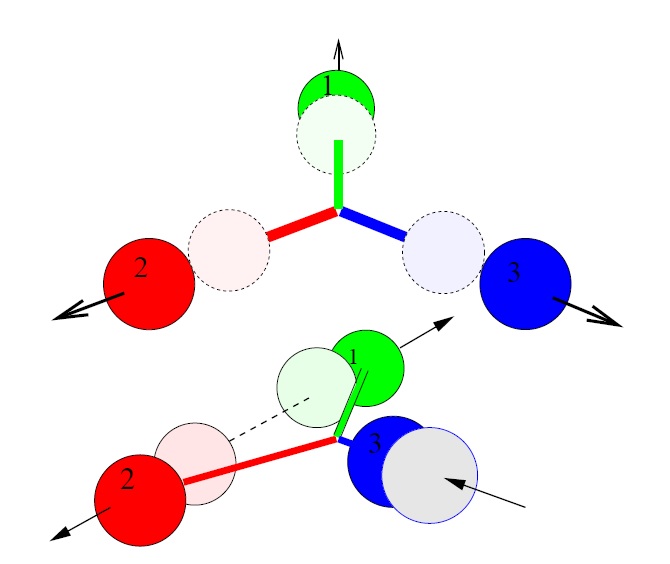}
\end{figure} 
Thus, the non-relativistic kinetic energy part for three identical particle system is \cite{Santopinto:2004hw}
\begin{equation}
	H = 3m + \frac{{\bf P^{2}_{\rho}}}{2m} + \frac{{\bf P^{2}_{\lambda}}}{2m} 
\end{equation}
where, $m=\frac{m_{\rho}m_{\lambda}}{m_{\rho}+m_{\lambda}}$ is the reduced mass.
\begin{equation}
	m_{\rho}=\frac{2m_1m_2}{m_1+m_2} 
	\hspace{0.5cm} and \hspace{0.5cm}
	m_{\lambda}=\frac{2m_3(m_{1}^{2}+m_{2}^{2}+m_{1}m_{2})}{(m_1+m_2)(m_1+m_2+m_3)}
\end{equation}
We introduce hyperspherical coordinates as hyperradius x and hyperangle $\xi$ from Jacobi coordinates as \cite{Bijker:1998zb},
\begin{equation}
	x = \sqrt{{\bf \rho^{2}} + {\bf \lambda^{2}}} ; \; \; \xi = arctan(\frac{\rho}{\lambda})
\end{equation}
The model requires that the potential to be chosen should be depending only on hyperradius x. However, hypercentral potential is not a pure two body interaction but can also contain three body terms.
The hyper-radial equation whose solution is $\psi(x)$ is as follows,
\begin{equation}
	\left[\frac{d^{2}}{dx^{2}} + \frac{5}{x}\frac{d}{dx} - \frac{\gamma(\gamma +4)}{x^{2}}\right]\psi(x) = -2m[E-V_{3q}(x)]\psi(x) 
\end{equation}
In the hypercentral approximation, the potential is expressed in terms of the hyperradius as \cite{Giannini:2015dca}
\begin{equation}
	\sum_{i<j} V(r_{ij})=V(x)+ ...
\end{equation}

\begin{figure}
	\centering
	\includegraphics[scale=0.26]{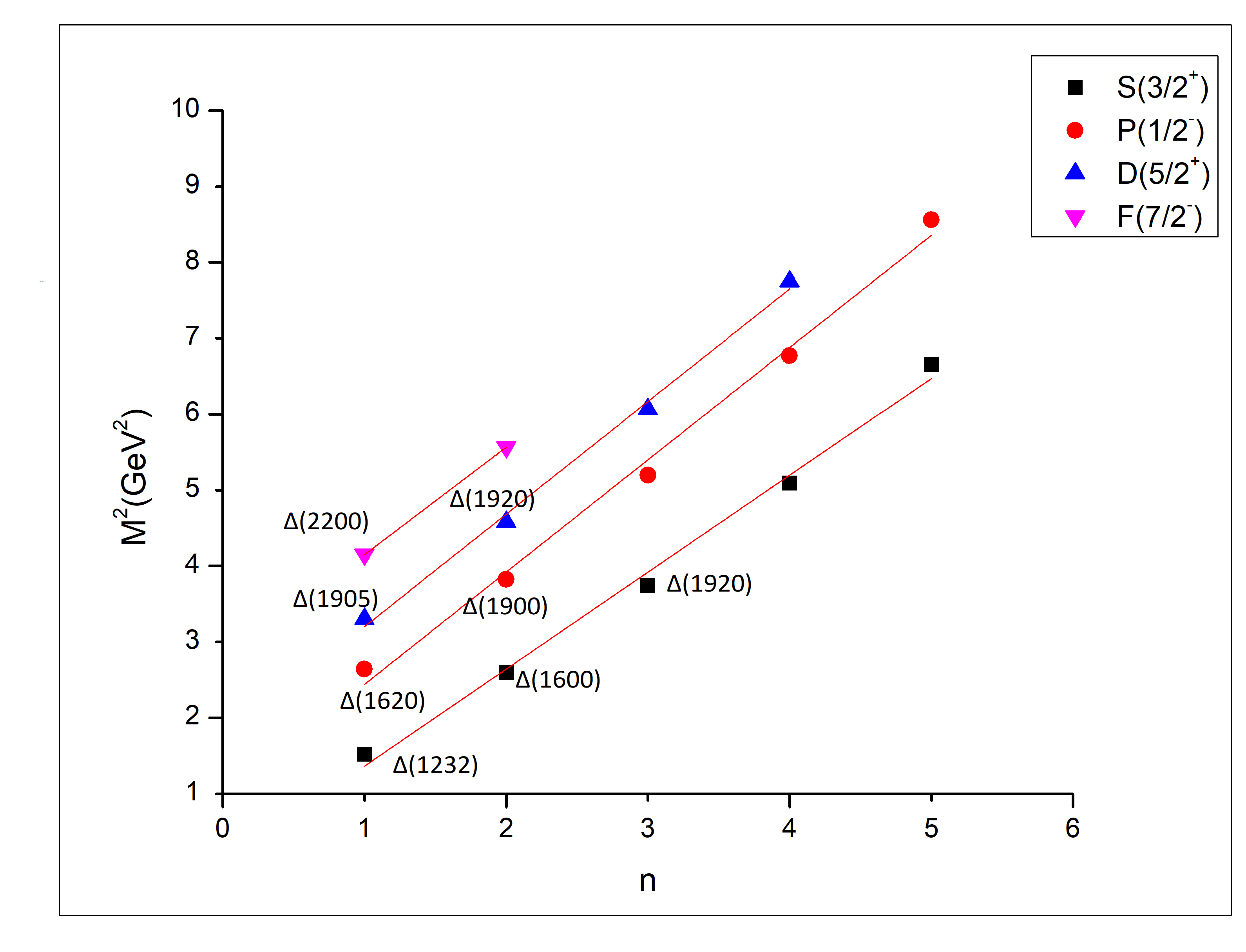}
	\caption{\label{fig:delta-n} (n,$M^{2}$) Regge trajectory for $\Delta$ states \cite{Menapara:2020dhy}}
\end{figure}

\begin{figure}
	\centering
	\includegraphics[scale=0.27]{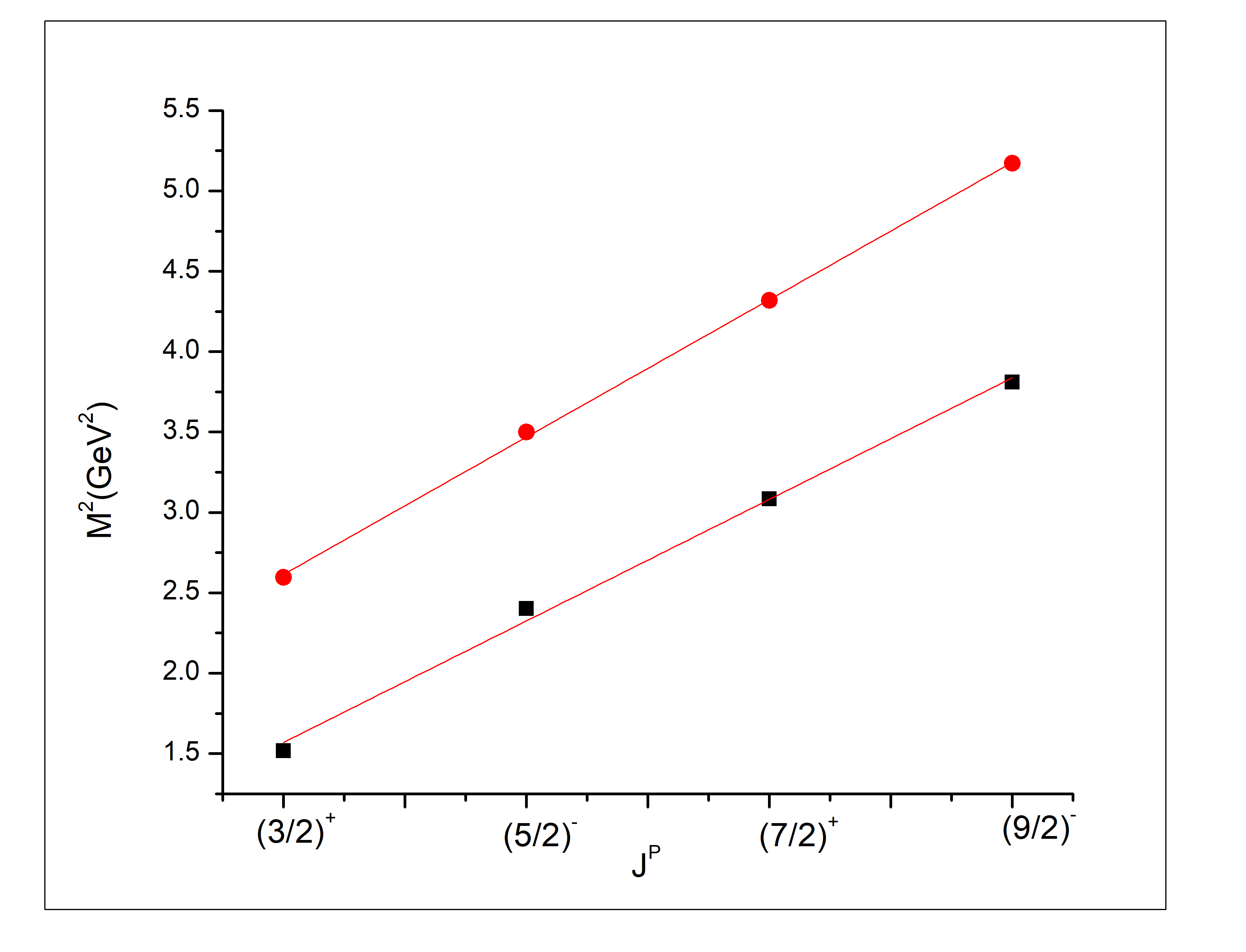}
	\caption{\label{fig:delta-j2} (J,$M^{2}$) Regge trajectory for $\Delta$ states \cite{Menapara:2022ksj}}
\end{figure}

%
%
%

In the present study, the deduced choice for the hypercentral potential is hypercoulomb type
\begin{equation}
	V(x) = -\frac{\tau}{x} 
\end{equation}
Here, the hyper-Coulomb strength $\tau=\frac{2}{3}\alpha_{s}$ where, $\frac{2}{3}$ is color factor for baryon. The negative sign represents the strong attractive interaction between quarks. The fundamental gauge theory of the strong interactions, quantum chromodynamics (QCD), uses the function $\alpha_{s}(Q^{2})$ to set the strength of interactions involving quarks and gluons as a function of momentum transfer Q. To describe hadronic interactions at both long and short distances, it is necessary to comprehend the behaviour and magnitude of the QCD coupling over the entire $Q^{2}$ range \cite{Deur:2016tte}. To test high-energy models and match the increasing accuracy of hadron scattering experiments at high $Q^{2}$ (short distances), precise knowledge of $\alpha_{s}(Q^{2})$ is required. It is also necessary to know the behavior at low $Q^{2}$ (long distances), such as the scale of the proton mass, in order to understand hadronic structure, quark confinement and hadronization processes.
\begin{equation}
	\alpha_s = \frac{\alpha_s(\mu_0)}{1+\left( \frac{33-2n_f}{12\pi}\right)\alpha_s(\mu_0)ln\left(\frac{m_1+m_2+m_3}{\mu_0}\right)}.
\end{equation}
Here, $\alpha_{s}$ is 0.6 at $\mu_{0}=1 GeV$. $n_f$ gives the number of active quark flavors which is 3 here \cite{Prosperi:2006hx}. The highest value for $n_f$ can be 6. 
The spin-dependent terms added are,
\begin{eqnarray}
	V_{SD}(x) = V_{LS}(x)({\bf L \cdot S}) + V_{SS}(x) \left[S(S+1)-\frac{3}{2}\right] \\ + V_{T}(x)\left[S(S+1) -\frac{3({\bf S \cdot x})({\bf S \cdot x})}{x^{2}} \right] 
\end{eqnarray}
Let, $V_{V}= \frac{\tau}{x} $ and $V_{S}= \alpha x $ \\
The spin-orbit term, 
\begin{equation}
	V_{LS}(x)= \frac{1}{2m_{\rho}m_{\lambda}x}\left( 3\frac{dV_{V}}{dx} - \frac{dV_{S}}{dx} \right) 
\end{equation}
The spin-spin term,
\begin{equation}
	V_{SS}(x) = \frac{1}{3m_{\rho}m_{\lambda}} \nabla^{2}V_{V}
\end{equation} 
The tensor term,
\begin{equation}
	V_{T}(x) = \frac{1}{6m_{\rho}m_{\lambda}} \left( \frac{d^{2}V_{V}}{dx^{2}}- \frac{1}{x} \frac{dV_{V}}{dx} \right) 
\end{equation}
{\bf L} and {\bf S} represent the angular momentum and total spin. 
a correction term with $\frac{1}{m}$ dependence is also considered as a refinement to the potential \cite{Brambilla:2000gk,Koma:2006si}.  
\begin{equation}
	V^{1}(x)= -C_{F}C_{A}\frac{\alpha_{s}^{2}}{4x^{2}}
\end{equation}
where $C_{F}$ and $C_{A}$ are Casimir elements of fundamental and adjoint representation with value $\frac{2}{3}$ and 3 respectively. It has been attempted to resolve the problem of hCQM having a lower mass for a higher spin state for a given angular momentum quantum number l by using a second-order correction of mass, which is written i.e. $\frac{1}{m^{2}}$ \cite{Pineda:2000sz,Chaturvedi:2019usm}. 
The O($\frac{1}{m^{2}}$) consists of spin-orbit and spin-tensor corrections described as \cite{koma1,Koma:2007jq},
\begin{equation}
	V_{ls}^{2}(x)= (\frac{c_{s}}{2x}\frac{dV}{dx} + \frac{c_{F}}{x}(V_{1}+V_{2})){{l\cdot s}}
\end{equation}
\begin{equation}
	V_{s12}^{2}(x)=\frac{c_{F}^{2}}{3}V_{3}{{s_{12}}} 
\end{equation}

\begin{equation}
	V_{1}=-(1-\epsilon)\alpha ;\quad V_{2}=\frac{\alpha^{'}}{x^{2}}+\epsilon\alpha ;\quad  V_{3}=\frac{3\sigma}{x^{3}}
\end{equation}
Here, $c_{F}=1+ \frac{\alpha_{s}}{2\pi}(C_{F}+C_{A})$ and $c_{s}=2c_{F}-1$ are the coefficients.
$\alpha^{'}=\frac{2}{3}\alpha_{s}$ and $\epsilon=0.2$ and $\sigma=0.216$ are the fitting parameters \cite{koma1}. Also, $V^2=V_{ls}^{2}(x) + V_{s12}^{2}(x)$. \\
The final Hamiltonian then becomes,
\begin{equation}
	H = \frac{P^{2}}{2m} + V(x) + V_{SD}(x) + \frac{1}{m}V^{1}(x) + \frac{1}{m^{2}}V^{2}(x)
\end{equation}
The constituent quark mass has been taken as $m_u=m_d=290 MeV$ and $m_s=500 MeV$ for u, d and s quark respectively.

\section{Resonance spectra of N to $\Omega$ baryons: A brief overview of results}

The resonance masses obtained with the above model has been subjected to comparison with various models whose details are presented in few of our articles. However, here recently added study with Regge phenomenology and well known relativistic model are compared with our data for strange baryons. Nucleons are shown in a generalized comparison in table  \ref{tab:nucleon}.

In case of N and $\Delta$ baryons, the number of one star states are less. The available four and three star states have been attempted to reproduce through the linear potential. It has been observed that low-lying states are well matched with experimental data as well as other theoretical models. But without correction the higher excited resonances are going towards over prediction. Here, figures \ref{fig:delta-n} and \ref {fig:delta-j2} represents the linear Regge trajectories for the spectra of $\Delta$ baryon.\\

For $\Lambda$ baryon, the more interesting part deals with associating 1 star stated resonances to our predictions. Also, 2S and 1P($\frac{3}{2}^{-}$) are within 20 MeV variation from PDG and other comparison. States 2000, 2050 and 2080 are a assigned to be members of 3P family with $\frac{1}{2}^{-}$, $\frac{3}{2}^{-}$ and $\frac{5}{2}^{-}$ respectively. The 1710 ($\frac{1}{2}^{+}$) is associated with 1D state but is comparatively lower than nearby spin-partners. 2070($\frac{3}{2}^{+}$) and 2085($\frac{7}{2}^{+}$) is perfectly fit to 2D$\frac{3}{2}^{+}$ and 2D$\frac{7}{2}^{+}$ respectively. Table \ref{tab:lambda} shows that few of the states are well matched with Regge results and relativistic ones. \\

For $\Sigma$ baryon, four star states are well established and so are the three star states  as shown in table \ref{tab:Sigma}. The two and one star states have been attempted to explore. The two star state 1880 might be somewhere in 1D family but in the present data nearly 120 MeV higher predicted. The 1580 state with ($\frac{3}{2}^{-}$) is not reproduced in the current study. Within 50 MeV apart 1780 could be 2S($\frac{3}{2}^{+}$). 1940 state best matches with 1D($\frac{3}{2}^{+}$) by 60 MeV difference. 2110 is compatible with 2P($\frac{1}{2}^{-}$) by 15 MeV. The 2230 state is exactly corresponding with 3S($\frac{3}{2}^{+}$). The 2455 MeV state has no known JP value but present study finds it the positive parity 2D family. The 3000 and 3170 could be 5S states with ($\frac{1}{2}^{+}$) and ($\frac{3}{2}^{+}$).\\

For limited experimental data of $\Xi$ our study has gathered a sizeable number of states. In comparison to 1950 of PDG, this model has produced 1971 and 1964 for $\frac{3}{2}^{+}$ as marked by table \ref{tab:XI}. The 1820 state also deviates by 30 MeV from experimental findings. The existing model, however, does not naturally generate the restricted states 1620 and 1690. Assigning $\frac{1}{2}^{-}$ to PDG state 2370, we see that it matches our 2P masses of 2373 and 2333 as well as our 3S state ($\frac{1}{2}^{-}$). The $\Xi$(2250) is very consistent with our 1D$\frac{5}{2}^{+}$ state. Additionally, the 2500 state agrees with our results' 3S($\frac{3}{2}^{+}$). Based on the information currently available, this assignment is correct, but true confirmation will depend on the precise experimental results in the next years. \\

With only the ground state totally known for $\Omega$, there are numerous possibilities to consider. The 1P($\frac{1}{2}^{-}$) state matches the $\Omega$(2012) state \ref{tab:Omega}. Additionally, the three star state 2250 agrees with the current model's 1D($\frac{3}{2}^{+}$), but it can also be given any of the positive parity J values for 1D. The two star state 2380 is the closest to our 2P, or negative parity, state, but the precise J value assignment is still awaiting, with values as wide as $\frac{1}{2}^{-}$, $\frac{3}{2}^{-}$ and $\frac{5}{2}^{-}$.\\

Theoretical and experimental research into the electromagnetic properties of baryons is a bustling field. This inherent quality sheds light on the form and other transitional dynamics of decay mode dynamics. The generalized form of magnetic moment is
\begin{equation}
	\mu_{B}= \sum_{q} \left\langle \phi_{sf} \vert \mu_{qz}\vert \phi_{sf} \right\rangle
\end{equation}
where $\phi_{sf}$ is the spin-flavour wave function. The contribution from individual quark appears as 
\begin{equation}
	\mu_{qz}= \frac{e_{q}}{2m_{q}^{eff}}\sigma_{qz}
\end{equation}
$e_{q}$ being the quark charge, $\sigma_{qz}$ being the spin orientation and $m_{q}^{eff}$ is the effective mass which may vary from model based quark mass due to interactions. Here, is it noteworthy that magnetic moment shall have contribution from many other effects within the baryon as sea quark, valence quark, orbital etc. Table \ref{tab:mm} and \ref{tab:radiative} here reflects the magnetic moment and radiative decay widths which have been obtained using the calculated resonances.
\begin{table*}
	\renewcommand{\arraystretch}{1.2}
	\centering
	\caption{\label{tab:mm} Magnetic moment for ground states of all strange baryons in addition with their final spin-flavour wave-function.}
	
	\begin{tabular}{ccccc}
		\hline
		Spin & Baryon & $\sigma_{qz}$ & Mass (MeV) & $\mu$ ($\mu_{N}$)\\
		\hline
		$\frac{1}{2}$ & $\Xi^{0}$(uss) & $\frac{1}{3}(4\mu_{s}-\mu_{u})$ & 1322 & -1.50 \\
		$\frac{1}{2}$ & $\Xi^{-}$(dss) & $\frac{1}{3}(4\mu_{s}-\mu_{d})$ & 1322 & -0.46\\
		$\frac{3}{2}$ & $\Xi^{*0}$(uss) & $(2\mu_{s}+\mu_{u})$ & 1531 & 0.766 \\
		$\frac{3}{2}$ & $\Xi^{*-}$(dss) & $(2\mu_{s}+\mu_{d})$ & 1531 & -1.962 \\
		$\frac{1}{2}$ & $\Sigma^{+}$(uus)& $\frac{1}{3}(4\mu_{u}-\mu_{s})$ & 1193 & 2.79\\
		$\frac{1}{2}$ & $\Sigma^{0}$(uds)& $\frac{1}{3}(2\mu_{u}+2\mu_{d}-\mu_{s})$ & 1193 & 0.839 \\
		$\frac{1}{2}$ & $\Sigma^{-}$(dds)& $\frac{1}{3}(4\mu_{d}-\mu_{s})$ & 1193 & -1.113 \\
		$\frac{3}{2}$ & $\Sigma^{*+}$(uus)& $(2\mu_{u}+\mu_{s})$ & 1384 & 2.877 \\
		$\frac{3}{2}$ & $\Sigma^{*0}$(uds)& $(\mu_{u}+\mu_{d}+\mu_{s})$ & 1384 & 0.353 \\
		$\frac{3}{2}$ & $\Sigma^{*-}$(dds)& $(2\mu_{d}+\mu_{s})$ & 1384 & -2.171 \\
		$\frac{1}{2}$ & $\Lambda^{0}$(uds)& $\mu_{s}$ & 1115 & -0.606 \\
		\hline
	\end{tabular}
\end{table*}

The detailed results for all these baryons without, with first order and second order correction and elaborated in context of all the possible models can be found in our articles \cite{Menapara:2020dhy,Menapara:2023rur,Menapara:2021dzi,Menapara:2021vug,Menapara:2023fgk}. Not only these, the magnetic moments, radiative decay widths and strong decay widths have also been looked into.
\begin{table*}
	\centering
	\caption{$\Sigma^{*}$ Radiative Decays}
	\label{tab:radiative}
	\begin{tabular}{cccc}
		\hline
		Decay & Wave-function & Transition moment(in $\mu_{N}$) & $\Gamma_{R}$(in MeV)\\
		\hline
		$\Sigma^{*0} \rightarrow \Lambda^{0}\gamma$ & $\frac{\sqrt{2}}{\sqrt{3}}(\mu_{u}-\mu_{d})$ & 2.296 & 0.4256 \\
		$\Sigma^{*0} \rightarrow \Sigma^{0}\gamma$ & $\frac{\sqrt{2}}{3}(\mu_{u}+\mu_{d}-2\mu_{s})$ & 0.923 & 0.0246 \\
		$\Sigma^{*+} \rightarrow \Sigma^{+}\gamma$ & $\frac{2\sqrt{2}}{3}(\mu_{u}-\mu_{s})$ & 2.204 & 0.1404 \\
		$\Sigma^{*-} \rightarrow \Sigma^{-}\gamma$ & $\frac{2\sqrt{2}}{3}(\mu_{d}-\mu_{s})$ & -0.359 & 0.0037 \\
		\hline
	\end{tabular}
\end{table*}

\section{Conclusion}
The present article reviewed the assumptions and findings of some major approaches towards the study of light baryon resonance spectra. A detailed comparison of the results shall allow us to improve the shortcomings as a single model hasn't predicted all the observed states precisely. This is also an opportunity to collate the experiment and theory predictions to look for yet unobserved states. Few of the models allow us to comment on the nature of a given state. Not only for resonance masses, but other baryon properties are also being studied with the basis of these models. Magnetic moment has been obtained for baryons. Availability of sizeable resonance states also draw us towards looking for all possible decay channels.  \\

The comparison of the results have been updated with new model predictions and the known relativistic approach. Regge phenomenology that started with the assumption of linearity of J and n plane curves has shown quite a good results. Also, applying the higher order corrections of mass to a non-relativistic model has lead to resolution of spin-hierarchy which wasn't achieved earlier using hCQM. Non-relativistic model as started since Capstick and Isgur has been updated with a every attempt to reproduce the observed spectra. This study is expected to be useful tool for experiments wherein new states are coming up and allowing us to identify them with other properties as well.     

Few highlights of overall article based on review of our previous results and comparison with other approaches are as follows:
\begin{itemize}
	\item In case of nucleon, low energy states have shown good agreement with experimental results and other models but current study also obtained higher spin states precisely for $\frac{9}{2}$ and $\frac{11}{2}$ which are also well in experimental range. 
	\item For $\Lambda$ baryon, the 1710 ($\frac{1}{2}^{+}$) is associated with 1D state but is comparatively lower than nearby spin-partners. 2070($\frac{3}{2}^{+}$) and 2085($\frac{7}{2}^{+}$) is perfectly fit to 2D$\frac{3}{2}^{+}$ and 2D$\frac{7}{2}^{+}$ respectively.
	\item This study has been able to comment on few of the states of $\Xi$  and $\Omega$ baryons where the availability of experimental data is scarce. The detailed results in our articles have shown a range of resonances upto hogher spin partners for upcoming experiments. 
\end{itemize}

\section*{Acknowledgement}
Ms. C. Menapara acknowledges the support from DST under INSPIRE FELLOWSHIP Scheme IF180733.

\end{document}